\documentclass[conference]{IEEEtran}
\IEEEoverridecommandlockouts
\usepackage{cite}
\usepackage{amsmath,amssymb,amsfonts}
\usepackage{algorithm}
\usepackage{algpseudocode}
\usepackage{graphicx}
\usepackage{siunitx}
\usepackage{textcomp}
\usepackage{multirow}
\usepackage{arydshln}
\usepackage{caption}
\usepackage{subcaption}
\usepackage{multirow}
\usepackage{booktabs}
\ifCLASSOPTIONcompsoc
\usepackage[caption=false,font=normalsize,labelfont=sf,textfont=sf]{subfig}
\else
\usepackage[caption=false,font=footnotesize]{subfig}
\fi
\usepackage{tikz}

\usepackage{acro}
\usepackage{xcolor}

\def\BibTeX{{\rm B\kern-.05em{\sc i\kern-.025em b}\kern-.08em T\kern-.1667em\lower.7ex\hbox{E}\kern-.125emX}}

\makeatletter
\let\old@ps@headings\ps@headings
\let\old@ps@IEEEtitlepagestyle\ps@IEEEtitlepagestyle
\def\confheader#1{%
\def\ps@headings{%
\old@ps@headings%
\def\@oddhead{\strut\hfill#1\hfill\strut}%
\def\@evenhead{\strut\hfill#1\hfill\strut}%
}%
\def\ps@IEEEtitlepagestyle{%
\old@ps@IEEEtitlepagestyle%
\def\@oddhead{\strut\hfill#1\hfill\strut}%
\def\@evenhead{\strut\hfill#1\hfill\strut}%
}%
\ps@headings%
}
\makeatother
\confheader{%
2024 14th International Conference on Indoor Positioning and Indoor Navigation (IPIN)
}

\DeclareAcronym{AI}{
    short = {AI},
    long = {Artificial Intelligence}
}

\DeclareAcronym{UWB}{
    short = {UWB},
    long = {ultra-wideband}
}

\DeclareAcronym{AOA}{
    short = {AOA},
    long = {angle-of-arrival}
}

\DeclareAcronym{MIMO-OFDM}{
    short = {MIMO-OFDM},
    long = {multiple-output orthogonal frequency-division multiplexing}
}

\DeclareAcronym{SISO}{
    short = {SISO},
    long = {single-input single-output}
}

\DeclareAcronym{TF}{
    short = {TF},
    long = {transformer}
}

\DeclareAcronym{FP}{
    short = {FP},
    long = {fingerprinting}
}

\DeclareAcronym{LOS}{
    short = {LOS},
    long = {line-of-sight}
}

\DeclareAcronym{TOA}{
    short = {TOA},
    long = {time-of-arrival}
}

\DeclareAcronym{NLOS}{
    short = {NLOS},
    long = {non-line-of-sight}
}

\DeclareAcronym{CIR}{
    short = {CIR},
    long = {channel impulse response}
}

\DeclareAcronym{VAE}{
    short = {VAE},
    long = {variational autoencoder}
}

\DeclareAcronym{DVAE}{
    short = {DVAE},
    long = {dynamical variational autoencoder}
}

\DeclareAcronym{DL}{
    short = {DL},
    long = {deep learning}
}

\DeclareAcronym{OOD}{
    short = {OOD},
    long = {out of distribution}
}

\DeclareAcronym{CNN}{
    short = {CNN},
    long = {convolutional neural network}
}

\DeclareAcronym{ML}{
    short = {ML},
    long = {machine learning}
}

\DeclareAcronym{SVM}{
    short = {SVM},
    long = {support vector machine}
}

\DeclareAcronym{VRNN}{
    short = {VRNN},
    long = {variational recurrent neural network}
}

\DeclareAcronym{ELBO}{
    short = {ELBO},
    long = {evidence lower bound}
}

\DeclareAcronym{TISSI}{
    short = {TISSI},
    long = {time index signal strength indicator}
}

\DeclareAcronym{LSTM}{
    short = {LSTM},
    long = {long short-term memory}
}

\DeclareAcronym{ROC}{
    short = {ROC},
    long = {receiver operating characteristic}
}

\DeclareAcronym{AUROC}{
    short = {AUROC},
    long = {area under receiver operating characteristic curve}
}

\DeclareAcronym{MAE}{
    short = {MAE},
    long = {mean absolute error}
}

\DeclareAcronym{MSE}{
    short = {MSE},
    long = {mean squared error}
}
\DeclareAcronym{TPR}{
    short = {TPR},
    long = {true positive rate}
}
\DeclareAcronym{TNR}{
    short = {TNR},
    long = {true negative rate}
}

\DeclareAcronym{RF}{
    short = {RF},
    long = {radio frequency}
}

\DeclareAcronym{ToA}{
    short = {ToA},
    long = {time-of-arrival}
}

\DeclareAcronym{SSM}{
    short = {SSM},
    long = {state space model}
}

\DeclareAcronym{EKF}{
    short = {EKF},
    long = {extended Kalman filter}
}

\DeclareAcronym{LiDAR}{
    short = {LiDAR},
    long = {light detection and ranging}
}

\DeclareAcronym{RX}{
    short = {RX},
    long = {radio receiver}
}

\DeclareAcronym{TX}{
    short = {TX},
    long = {radio transmitter}
}

\DeclareAcronym{GPT}{
    short = {GPT},
    long = {generative pre-trained transformer}
}

\title{Radio Foundation Models: Pre-training Transformers for 5G-based Indoor Localization}

\def\authorrefmark#1{\ensuremath{^{\textbf{#1}}}}
\author{
    Jonathan Ott\authorrefmark{1},
    Jonas Pirkl\authorrefmark{1},
    Maximilian Stahlke\authorrefmark{1,2},
    Tobias Feigl\authorrefmark{1,2},
    and Christopher Mutschler\authorrefmark{1}\\
    {\tt\footnotesize\{jonathan.ott, jonas.pirkl, maximilian.stahlke, tobias.feigl, christopher.mutschler\}@iis.fraunhofer.de}\vspace{+1mm}\\ 
    \IEEEauthorblockA{\authorrefmark{1}
        Fraunhofer Institute for Integrated Circuits (IIS), 
        Division Positioning and Networks, 
        90411 Nürnberg, Germany
        }
    \IEEEauthorblockA{\authorrefmark{2}
        Faculty of Engineering,
        Friedrich-Alexander-Universität Erlangen-Nürnberg, 
        91052 Erlangen, Germany 
    }
}

\begin{document}
\maketitle

\begin{abstract}
\ac{AI}-based radio \ac{FP} outperforms classic localization methods in propagation environments with strong multipath effects.
However, the model and data orchestration of \ac{FP} are time-consuming and costly, as it requires many reference positions and extensive measurement campaigns for each environment.
Instead, modern unsupervised and self-supervised learning schemes require less reference data for localization, but either their accuracy is low or they require additional sensor information, rendering them impractical.

In this paper we propose a self-supervised learning framework that pre-trains a general \ac{TF} neural network on 5G channel measurements that we collect on-the-fly without expensive equipment.
Our novel pretext task randomly masks and drops input information to learn to reconstruct it.
So, it implicitly learns the spatiotemporal patterns and information of the propagation environment that enable \ac{FP}-based localization.
Most interestingly, when we optimize this pre-trained model for localization in a given environment, it achieves the accuracy of state-of-the-art methods but requires ten times less reference data and significantly reduces the time from training to operation.
\end{abstract}

\vspace{+0.2cm}

\section{Introduction}

Accurate and robust radio-based localization becomes increasingly important for many emerging applications such as smart factories, healthcare, and emergency services~\cite{Laoudias2018}.
Classic approaches extract properties such as the \ac{TOA} of radio signals, determine distances between target and anchor nodes, and then estimate the position using multilateration~\cite{Dardari2015}.
Their major disadvantage is the need for \ac{LOS} to at least three anchor nodes.
However, due to the complexity of indoor environments, the \ac{LOS} is often blocked. 
This causes errors in the \ac{TOA} estimation and consequently also in localization. 
In complex environments \ac{AI}-based \ac{FP} enables localization. 
\ac{FP} exploits the fact that the propagation environment generates specific patterns in the received \acp{CIR} that are virtually unique for a specific position~\cite{Niitsoo2019}.
However, \ac{FP} requires large amounts of reference data, e.g., synchronized \acp{CIR} and associated reference positions at every potential position in space and for each new environment. 
Such data acquisition is costly and renders \ac{FP} impractical in real world scenarios.

A common, established approach that reduces the data acquisition effort for \ac{FP} is to pre-train it on synthetic data and then fine-tune it on little reference data of the real target environment~\cite{Stahlke2022}.
However, as synthetic data does not fully reflect reality, this approach still requires a lot of site-specific reference data, even for fine-tuning in the event of environmental changes.
To avoid reference data at all, others employ un- and self-supervised learning~\cite{Stahlke2023a, Stephan2024}.
However, they either require additional sensor information or their accuracy is significantly lower than that of \ac{FP}.
Self-supervised pre-training, such as \acp{GPT}~\cite{achiam2023gpt, touvron2023llama}, showed great success in natural language processing. 
Recent research, successfully adapted the concept to computer vision~\cite{Assran2023, Zhou2021} and time series~\cite{Baevski2020, Meng2023}. 
Their idea is to autonomously find and exploit implicit patterns of correlations and spatiotemporal relationships in large amounts of unlabeled data.
A predefined pretext task helps to learn meaningful representations from the data to avoid reference data.
The key lies in designing the pretext task so that the learned pattern can be reused in downstream tasks~\cite{Liu2021}.
There also exist self-supervised learning approaches for radio-signals. However, they rely on assumptions about the 
signal properties or additional spatial information \cite{Salihu2024}.

\begin{figure}[t!]
    \centering
    \includegraphics[width=\columnwidth,trim=0 0 0 0, clip]{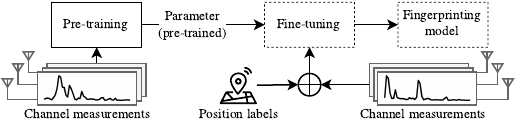}
    \vspace{-0.6cm}
    \caption{Two-step training pipeline: pre-training on \acp{CIR} wo. references (left) and fine-tuning on \acp{CIR} w. references (right).}
    \vspace{-0.3cm}
    \label{fig:pipeline}
\end{figure}

In this paper we propose a self-supervised learning framework with a novel pretext task to pre-train a \ac{TF}~\cite{Vaswani2017} neural network for radio \ac{FP}-based localization in a given environment without reference data.
Unlike other pretext tasks that learn general channel representations~\cite{Salihu2024}, we extract location-specific channel patterns from unlabeled sets of synchronized \acp{CIR} that we can freely collect, e.g., by crowdsourcing with mobile phones.
We define our pretext task to randomly mask and discard input information to learn to reconstruct it.
In this way, our predictive pretext task learns spatiotemporal patterns in \acp{CIR} that support downstream \ac{FP}-based localization.
In contrast to current approaches \cite{Salihu2024}, we do not augment the data by changing the signal shape. 
Hence we avoid inducing a bias in the learning procedure.
We then optimize the pre-trained \ac{TF} model on very few reference measurements from a real-world target environment to achieve accurate localization.
In real-world experiments with two 5G datasets, we show that our framework outperforms state-of-the-art supervised pre-training and \ac{FP} methods while requiring the least reference data.
In doing so, we move in the direction of generative pre-trained transformers for radio signals (RadioGPT) and show how \ac{GPT} may, for example, enable cost-effective, accurate and robust 5G-based indoor localization.

We structure the paper as follows.  
Sec.~\ref{sec:related_work} reviews related work. 
Sec.~\ref{sec:method} introduces our method.
Sec.~\ref{sec:experimental_setup} describes our experiments.
Sec.~\ref{sec:evaluation} discusses our results. 
Sec.~\ref{sec:conclusion} concludes.

\section{Related Work}\label{sec:related_work}

First, we discuss supervised \ac{FP} and pre-training schemes for radio-based localization.
We then discuss self-supervised learning schemes in general and for radio signaling.

The performance of classic localization approaches, such as \ac{TOA}- or \acl{AOA}-based methods, degrade in environments with strong multipath effects~\cite{Dardari2015}.
Deep learning-based radio \ac{FP} outperforms these approaches as it exploits the additional information provided by multipath to map synchronized \acp{CIR} to unique positions~\cite{Niitsoo2019}.
However, \ac{FP} is inaccurate and unreliable under environmental changes and so needs to be continuously updated.
This life-cycle management (initial training and updating) requires large amounts of labeled reference data, which is laborious and costly to acquire, rendering \ac{FP} impractical.
Thus, Sousa~\textit{et al.}~\cite{N.deSousa2018} pre-train \ac{FP} on synthetic data.
Others~\cite{Stahlke2022, Klus2021} use transfer learning to capture environment-specific~\cite{Stahlke2022} or signal-specific features~\cite{Klus2021} in pre-training and require fewer reference data of new environments to adapt and update.
However, they still require a significant amount of reference data.

Instead, self-supervised channel charting employs manifold learning that captures local geometries between \acp{CIR} using a distance metric without reference measurements.
The learned charts in combination with a few reference coordinates or other sensor information, e.g., directed velocities from inertial sensors, enable \ac{FP}-like localization.
However, without additional information, these approaches achieve lower localization accuracy than supervised \ac{FP}~\cite{Stahlke2023}.

Recently, self-supervised representation learning achieved great success in various application domains~\cite{Radford2018, Devlin2018, Assran2023, Zhang2023, Baevski2020}.
In general, it represents many different paradigms.
The most promising one exploits predictive pretext tasks during training by reconstructing information from a corrupted input (randomly masked and discarded)~\cite{Meng2023}.
Recently, Salihu~\textit{et al.}~\cite{Salihu2024} proposed a self-supervised learning framework for massive \ac{MIMO-OFDM} systems to learn frequency domain representations.
Their joint embedding approach learns to ignore stochastic extensions induced in the signals.
Although their approach is promising, it induces extensions that affect the spatial pattern in the data, such as random gains and phase changes, that render environment-specific optimization difficult and prevent localization.
Instead, we learn environment-specific representations in the time domain from sets of \acp{CIR} of a distributed \ac{SISO} topology.


\section{Methodology}\label{sec:method}

Our self-supervised \ac{FP} localization framework trains a deep neural network in two-stages (see Fig.~\ref{fig:pipeline}). 
First, it pre-trains the \ac{TF} using a self-supervised reconstruction task on unlabeled data to extract location-specific features from \acp{CIR}, i.e., synchronized \acp{CIR} of all base stations at a time or position.
Second, it fine-tunes the pre-trained \ac{TF} in a supervised manner on \acp{CIR} and reference positions to enable \ac{FP}.

\begin{figure}[b!]
    \centering
    \vspace{-0.7cm}
    \includegraphics[width=\columnwidth, trim=0 10 10 0, clip]{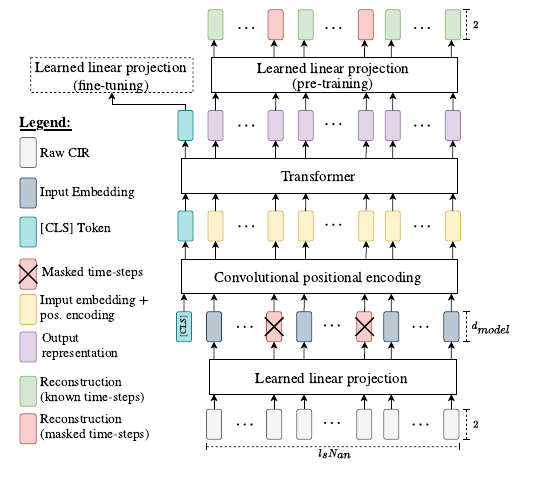}
    \vspace{-0.6cm}
    \caption{Schematic overview of our \ac{TF} pre-training method.}
    \label{fig:overview_method}
\end{figure}

\subsection{Transformer Neural Network: Refinements}\label{sec:meth:model}

Instead of a \ac{CNN} for \ac{FP}, we use a \ac{TF}~\cite{Vaswani2017} to learn spatiotemporal correlations of the components of the \acp{CIR}.
Fig.~\ref{fig:overview_method} shows our \ac{TF}.
We subsequently feed the representations (per-timestep or global) that \ac{TF} provides into a task specific layer (pretext or downstream) that generates the final predictions (see Fig.~\ref{fig:overview_method} last layer).

\subsubsection{Inputs}

On 1D sequences of high-dimensional vectors our \ac{TF} captures dependencies of the \acp{CIR} signal components between time steps across all antennas using the attention mechanism.
This is key to our pretext task of reconstructing masked signals.
To do this, we need to prepare the input data in two steps:
First, we align all \acp{CIR} of the training data to a sequence of two dimensional vectors (real and imaginary parts of the complex \ac{CIR} signal) of length $N_{an}l_s$ (gray boxes in Fig.~\ref{fig:overview_method}), where $N_{an}$ is the number of anchors and $l_s$ is the number of time steps per \ac{CIR}.
Next, we convert the 2D channel measurements into high-dimensional latent vectors of $d_{model}$ dimensions (dark blue boxes in Fig.~\ref{fig:overview_method}) via a {linear projection} with learnable parameters to match the size $d_{model}$ of the internal latent vectors of \ac{TF}. 
The high-dimensional input reduces the complexity of the discrimination and improves the flexibility for subsequent processing steps.

\subsubsection{Positional Encoding}

Unlike \acp{CNN} or recurrent neural networks, \ac{TF} treats each input time step individually.
Consequently, it does not interpret and exploit the order of the input sequence~\cite{Vaswani2017}.
To establish an internal order and to extract spatiotemporal information from \acp{CIR} with \ac{TF}, a convolutional layer~\cite{Baevski2020} impinges each data point with a learned positional encoding~\cite{Mohamed2019}. 
Instead of the lower attention layers, the convolutions also capture local features, that results in stable training.
The implicitly learned positional encoding enables the subsequent \ac{TF} layers to recover the sequence of the input.
We then feed the input embedding (with added positional encoding) into \ac{TF} (yellow blocks in Fig.~\ref{fig:overview_method}).

\subsubsection{Encoder based transformer} 

In contrast to vanilla encoder-decoder \acp{TF}, we implement an encoder \ac{TF}~\cite{Devlin2018} as we do not need the autoregressive generation of output representations during inference due to our fixed input and output sequence length.
Thus, information flows unconstrained bidirectionally along the input sequence. 
The model comprises $N_b$ blocks consisting of a multi-head attention layer with $h$ heads, followed by a positional feedforward network~\cite{Vaswani2017} with $d_{f\!f}$ hidden dimensions.
We use dropout for regularization with a rate of $P_{drop}$$=$0.2.

\subsection{Pretext Task - Reconstruction of \acp{CIR}}\label{sec:meth:masking}

To learn representations from \acp{CIR}, we use a predictive pretext task~\cite{Park2019} that learns to reconstruct masked, removed signal parts.
The idea is that the reconstruction of (environment-dependent) missing parts forces the model to learn spatiotemporal correlations. 
These are key for \ac{FP}-based localization.

A learned linear projection as the final layer maps the high-dimensional output representations of \ac{TF} to the 2D \ac{CIR} space (i.e., complex and imaginary parts).
Fig.~\ref{fig:overview_method} shows the known (green blocks) and the reconstructed signal parts (red blocks).
We compute the {loss} of the reconstruction task in the frequency domain.
Our experiments showed that this stabilizes the training process, as it is less sensitive to small time shifts in the \ac{CIR} than loss functions in the time domain.
We transform \acp{CIR} to the frequency domain via a differentiable fast Fourier transform.
Then, we compute the \ac{MSE} over the spectral components.

The masking is done immediately after we project the signal components into the high-dimensional latent space.
Our masking strategy replaces the hidden input time steps with a learned mask embedding (red boxes with black cross in Fig.~\ref{fig:overview_method}).
For each input signal, we generate $N_m$ masks.
Each mask comprises $l_m$ consecutive time steps and may overlap.
We define the time steps of a single mask such that $\{t_{m,0}, \dots, t_{m,0+l_m}\}$.
The first index of each mask $t_{m,0}$ is drawn from a uniform distribution such that $t_{m,0} \in [0, N_r - l_{m})$.
The upper bound of generated masks is determined by $N_m = P_m N_r / l_m$. 
$P_m \in [0, 1]$ describes the fraction of input time steps that are masked.
$P_m$ increases the (total) number of masked time steps, the probability of (partially) overlapping masks, the reconstruction effort, and thus the computational and learning effort.
For stable training, we linearly increase $P_m$ every $N_p$ epochs by 1 \% (in $P_{min}$ and $P_{max}$).

\begin{figure}
    \centering
    \includegraphics[width=\columnwidth]{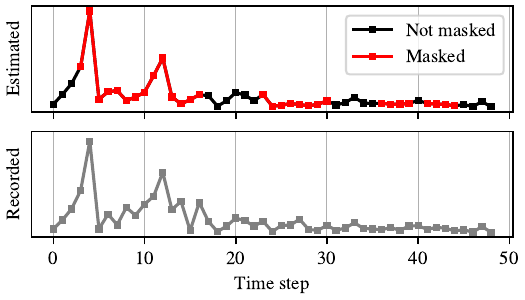}
    \vspace{-0.6cm}
    \caption{\ac{CIR}-magnitude of a fingerprint with $N_{an}$ \acp{CIR}. 
    Input \ac{CIR} (bottom) and reconstruction (top; red: masked parts).}
    \vspace{-0.5cm}
    \label{fig:reconstruction}
\end{figure}

Fig.~\ref{fig:reconstruction} shows a reconstruction of a single (red) masked \ac{CIR}.
Note that, only by pre-learned correlations of environment-specific information, our \ac{TF} can reconstruct the \ac{LOS} path (time step 4) and another multipath component (time step 12).

\subsection{Downstream Task: \ac{FP}-based Localization}\label{sec:meth:fingerprinting}

\ac{TF} generates a sequence of output representations with the same length as the input sequence.
For the \ac{FP} task, however, we do not generate a prediction for each time step, but map the entire channel measurement (fingerprint of \acp{CIR}) to a corresponding 2D reference position.
To do this, we prefix the input time series with a learnable [CLS] $\boldsymbol{E}_{cls}$ token (cyan box in Fig.~\ref{fig:overview_method})~\cite{Devlin2018}.
This instructs \ac{TF} to collect global information at this marked position, which serves as the global fingerprint representation.
Finally, we apply a learned linear projection to the output representation at the position of the [CLS] token, 
that maps the representation to the local two-dimensional coordinate system.
Note that parameters of this final layer and the final layer of the pretext task differ.

\section{Experimental Setup}\label{sec:experimental_setup}

Sec.~\ref{sec:exp:configuration} describes the model configurations.
Sec.~\ref{sec:exp:datasets} reports datasets. 
Sec.~\ref{sec:exp:baseline_metric} introduces evaluation metrics.

\subsection{Model configurations}\label{sec:exp:configuration}

We report our model configuration in 2 phases (pre-training, localization) and the baseline configuration.

\subsubsection{Pre-training phase}

Our \ac{TF} consists of $N_b$$=$4 encoder blocks.
Each block consists of a multi-head attention layer with $N_h$$=$16 heads, followed by a positional feedforward network with a hidden size of $N_{f\!f}$$=$1024.
The input embedding size is $d_{model}$$=$512.
Our \ac{TF} with 15 million trainable parameters optimizes computational effort and model size~\cite{Devlin2018}.
The masking parameters $P_{min}$$=$0.3, $P_{max}$$=$0.5, and $N_p$$=$50 optimize the objectives of the pretext task and enable stable training and high \ac{FP} localization accuracy.
For a fair comparison, we pre-train each \ac{TF} model for 192~\si{h} on 4 Tesla V100 graphics cards.
Note that \ac{TF} continues to optimize even beyond this training time.
Instead of a classical Adam optimizer, we use RAdam~\cite{Liu2019a} to avoid warm-up phase of the learning rate to stabilize training.

\subsubsection{Localization phase}

We initialize our \ac{TF} encoder with pre-trained weights conditional on two experiments:
(1) Pre-training and fine-tuning on data from the same environment (hereafter called TF-PT); 
(2) Pre-training on data from one of the two environments and fine-tuning on data from the other environments (hereinafter called TF-C-PT).
We define the loss function of the \ac{FP} localization as the squared Euclidean distance between the estimated and the reference position.
Early stopping ends the fine-tuning at convergence.

\subsubsection{Baseline models}

We compare our TF-PT and TF-C-PT with the methods of Stahlke~\textit{et~al.}~\cite{Stahlke2022} (CNN-PT) and Ott~\textit{et~al.}~\cite{Ott2023} (TF-SC). 
In contrast to TF-PT and TF-C-PT, we train CNN-PT and TF-SC in a supervised manner for direct \ac{FP}-based localization, regression.
CNN-PT employs a \ac{CNN} that we pre-train on labeled synthetic data and fine-tune on labeled real data.
This reduces the labeling effort. 
TF-SC employs a \ac{TF} that wa train from scratch on labeled real data.

\subsection{Datasets}\label{sec:exp:datasets}

We evaluate our self-supervised learning framework on two 5G datasets with different propagation environments.

\subsubsection{Real-World Datasets}

The two real-world datasets contain \ac{CIR} fingerprints of a 5G-compatible software defined radio setup (bandwidth: 100~\si{MHz}; carrier frequency: 3.75~\si{GHz}; reference signal: DL PRS \cite{dwivedi2021positioning}) and reference positions in the Fraunhofer test center in Nuremberg, Germany (area: about 40 $\times$ 30~\si{m}). 
We attach six synchronized transmitter antennas to the interior walls of the test center and record fingerprints (i.e., received \acp{CIR} of a signal burst from the transmitter) at 100~\si{Hz}.
An optical Nikon iGPS reference system provides accurate reference positions in the millimeter range for each receiver position in space.
The high recording frequency (100~\si{Hz}) of the fingerprints results in subsequent \acp{CIR} of similar signal pattern as the receiver moved slowly (human gait).
To ensure that we evaluate the (generalization) performance of all methods, we train and test on different trajectories. 

We consider two different propagation scenarios:
First, a scenario that mimics a realistic industrial environment with forklift and high shelves.
Second, a scenario including a narrow corridor and large walls that cause deterministic signal blockages and dense multipath.
For each scenario, we use about 280,000 fingerprints for training and about 70,000 for validation.
Note that this is only for pre-training.
For fine-tuning, we limit the training data.
The test sets consist of about 230,000 signal bursts.


\subsubsection{Synthetic Dataset}

It contains \acp{CIR}, that we generate with the geometry-based stochastic radio channel generator QuaDRiGa~\cite{Jaeckel2017} (bandwidth: 100~\si{MHz}; carrier frequency: 3.75~\si{GHz}) and reference positions in an indoor LoS scenario that mimics the environment of the real-world datasets (area: 27 $\times$ 15~\si{m}).
Again, we place six transmitters in the room.

The propagation scenario is parameterized according to 3GPP 38.901 InF-LoS specifications~\cite{3GPP2020}.
We simulate a total of 300,000 fingerprints at random positions and divide them into 48\% training, 12\% validation, and 40\% test data.
Note that we only use synthetic data to pre-train the CNN-PT.

\subsection{Evaluation metrics}\label{sec:exp:baseline_metric}

To quantify the positioning accuracy (i.e., error), we consider the Euclidean distance between estimated positions (2D coordinates) and reference positions.
We also determine the mean and the 90th percentile of the cumulative distribution function (CE90).
And we evaluate the effects of increasing amounts $N$ of training data on the \ac{FP} localization accuracy.

\section{Evaluation}\label{sec:evaluation}

\begin{figure*}[tb]
    \centering
    \begin{subfigure}[b]{0.99\columnwidth}
        \centering
        \includegraphics[width=\columnwidth, trim= 0 0 0 0, clip]{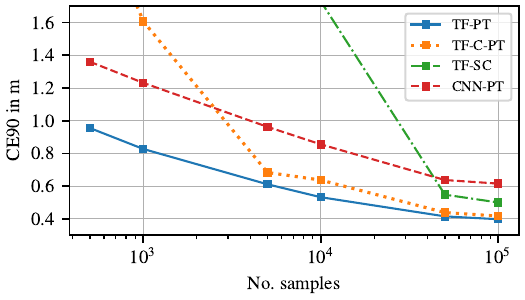}
        \vspace{-0.5cm}
        \caption{Results of unseen test data from the Industrial site.}
        \label{fig:ce90_mae}
    \end{subfigure}
    \begin{subfigure}[b]{0.99\columnwidth}
        \centering
        \includegraphics[width=\columnwidth, trim= 0 0 0 0, clip]{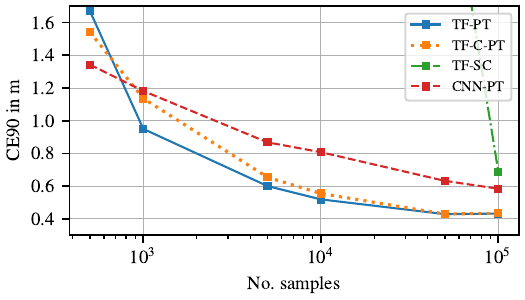}
        \vspace{-0.5cm}
        \caption{Results of unseen test data from the Deterministic site.}
        \label{fig:ce90_lshape}
    \end{subfigure}
    \vspace{-0.05cm}
    \caption{Localization error (CE90) of all methods w.r.t. the reference measurements we use for fine-tuning. 
    Trained and tested on data of the same scenario. Except TF-C-PT, that we pre-train on data of one scenario and test it on data of the other scenario.}
    \label{fig:errorgraphs}
    \vspace{-0.6cm}
\end{figure*}

Sec.~\ref{sec:eval:specific} discusses results of pre-training and fine-tuning in the same scenario.
Sec.~\ref{sec:eval:generic} discusses results of pre-training in one scenario and fine-tuning in the other.
Note that the synthetic dataset is only required to pre-train the baseline CNN-PT~\cite{Stahlke2022}, as the approach requires labeled data.
For pre-training all other methods, we only use unlabeled samples of the real datasets.

\subsection{Environment-Specific Evaluation}
\label{sec:eval:specific}

Tbl.~\ref{tab:results} lists all results, the best results are bold. 
In general, our novel TF-PT (almost) always localizes most accurately in both scenarios regardless of $N$ reference measurements (see Figs.~\ref{fig:ce90_mae} and \ref{fig:ce90_lshape}).

TF-PT achieves the highest localization accuracy of all methods along all configurations already on $N$$=$10k reference measurements on both datasets: 0.532~\si{m} in the Industrial dataset and 0.519~\si{m} in the Deterministic dataset. 
The baseline CNN-PT requires 10 times more data samples to achieve similar results: 0.616~\si{m} for the Industrial and 0.584~\si{m} for the Deterministic dataset. 


It is interesting that TF-PT localizes slightly less accurate (1.67~m) than most other approaches on the Deterministic dataset at $N$$=$500. 
We think this is due to the spatial distribution of the unsupervised dataset. 
The deterministic environment has a higher spatial density of samples on the right hand side for the training dataset, while the left hand side is very sparse. 
Thus, the low number of random samples does not cover the data on the left hand side very well, which leads to lower accuracies in the fine-tuning. 
This assumption explains also the poor performance of our model trained on the Deterministic dataset (TF-C-PT) and fine-tuned on the Industrial dataset, as the left hand side, despite the fact of a different composition of the environment, is underrepresented. 
In contrast, the baseline CNN-PT employed uniformly distributed samples across the environment, which leads to a better spatial distribution for small datasets.

Given a large number of data samples, TF-PT localizes more accurate, from $N$$=$500 with a CE90 = 0.955~\si{m} to $N$$=$100k with a CE90 = 0.398~\si{m}, than all other methods on the Industrial dataset with the same configuration.
We believe that our novel pre-training strategy forces the \ac{TF} to model the environmental information of the environment intrinsically without any specific task. 
Downstream tasks that explicitly exploit this environmental information, such as fingerprinting, can therefore highly benefit from the learned intrinsic spatial structure. 
Thus, with only few training samples, our model learned to efficiently interpolate, which leads to a high performance. 
Evaluations on the \ac{TF} trained without the pre-training strategy, emphasize the importance as the accuracy is very low across all datasets and data set sizes, shown in Figs.~\ref{fig:ce90_mae} and \ref{fig:ce90_lshape} as TF-SC. 
Thus, the \ac{TF} it-self does not enable the high accuracy, but the pre-training strategy along with the efficiency of the \ac{TF}. 
In contrast, the baseline CNN-PT learned a fingerprinting model on a simplified dataset. 
Thus, the model learned less realistic features for the downstream task as it may have overfitted to the simplified dataset.

\begin{table}[tb]
    \centering
    \caption{Localization error (CE90 in~\si{m})}
    \begin{tabular}{c|c|c|c|c|c}
        & N & TF-PT & TF-C-PT & TF-SC & CNN-PT \\ \hline\hline
        \multirow{4}{*}{Industrial}  & 500 & \textbf{0.955} & 2.401 & - & 1.361 \\ 
        & 1k & \textbf{0.827} & 1.608 & - & 1.231 \\ 
        & 10k & \textbf{0.532} & 0.637 & 1.730 & 0.855 \\ 
        & 100k & \textbf{0.398} & 0.417 & 0.500 & 0.616 \\ \hline\hline
        \multirow{4}{*}{Deterministic} & 500 & 1.671 & 1.542 & - & \textbf{1.343} \\ 
        & 1k & \textbf{0.950} & 1.136 & - & 1.182 \\ 
        & 10k & \textbf{0.519} & 0.555 & 2.707 & 0.808 \\ 
        & 100k & \textbf{0.432} & 0.435 & 0.688 & 0.584 \\
    \end{tabular}
    \label{tab:results}
\end{table}

\subsection{Environment-Independent Evaluation}
\label{sec:eval:generic}

In this evaluation, we want to evaluate the effect of site-specific and deployment system specific pre-training. 
Thus, we compare our model pre-trained and fine-tuned on data of the same site (TF-PT) and pre-trained on one site and fine-tuned on the other site (TF-C-PT). 
The cross-site results are listed in Tbl.~\ref{tab:results} (TF-C-PT) and as orange curves in Figs.~\ref{fig:ce90_mae} and \ref{fig:ce90_lshape}. 
In general we can see that the \ac{TF} also benefits from unlabeled data from a different site, but with a similar deployment. 
In the Industrial site, the \ac{TF} pre-trained in the same environment (TF-PT) always outperforms the \ac{TF} pre-trained in the Deterministic site (TF-C-PT). 
However, given $N$$=$100k samples, TF-C-PT approaches the accuracy of the site-specific \ac{TF}, TF-PT, with only a difference of 0.03~\si{m} in the CE90. 
We can see a very similar behavior in the Industrial scene. 
However, for fewer number of samples the performance of the TF-PT and TF-C-PT diverges. 
This can be explained by the non-uniform data distribution from the deterministic environment used to pre-train TF-C-PT, as already elaborated in Sec.~\ref{sec:eval:specific}. 
Thus, more labeled samples are required to overcome the holes of the Deterministic data distribution.

In general, our evaluations have shown that our pre-training strategy is not limited to site specific data, but also benefits from deployment specific information from a different environment. 
This renders our approach flexible w.r.t. the availability of data, render our approach a viable candidate for a foundation model for fingerprinting.





\section{Discussion}

Next, we discuss the limitations of our approach and possible directions for research to address these limitations.
While the choice of \ac{TF} as neural network backbone leads to excellent results, training it requires a lot of computational resources.
In addition, \ac{TF} suffers from long inference times compared to CNN-based approaches.
This is due to the large amount of trainable parameters and the attention mechanism that has a quadratic increasing complexity with the length of the input sequence. 
As an alternative to the \ac{TF} model, selective state-space-models have shown promising results with time-series data such as audio recordings.
The MAMBA model~\cite{Gu2024} achieves a significant speedup in the inference and scales linearly in runtime with the sequence length. 
This is especially interesting for radio localization and communication, as inference time is a crucial factor due to high recording frequencies and high demands on real-time capabilities. 

Our evaluations have also shown that site-specific data for pre-training is superior as the \ac{TF} may encode environment specific information. 
However, deployment specific data, i.e., data in a different environment but with a similar radio setup, also helped to lower the required number of labeled samples. 
While a foundation model might ideally be general to site and radio setup, more specific datasets for pre-training improves the foundation for the downstream task. 
Thus, our approach renders flexibility to w.r.t. data availability.

Another important aspect is the spatial coverage of the pre-training data. 
Our evaluations have shown that fine-tuning on a pre-trained model worked best if the unsupervised dataset covers the target downstream fingerprinting task. 
While we do not provide positional information during the self-supervised pre-training, the information is still intrinsically encoded in the manifold of the data~\cite{Stahlke2023}. 
Thus, fingerprint based foundation models may cover the same area as the downstream task if possible to achieve the highest efficiency. 



\section{Conclusion}\label{sec:conclusion}

In this paper, we propose a self-supervised pre-training method for a \ac{TF} neural network for radio \ac{FP}-based localization, serving as a foundation model for radio fingerprinting. 
It intrinsically learns spatiotemporal patterns from \ac{FP} channel measurements of a propagation environment without any reference information. 
We formulate a novel pretext task, i.e., masking fractions of \acp{CIR}s, that forces a \ac{TF} to encode environment specific information from the channel information, which improves the fine-tuning performance of a environment specific downstream task, i.e., fingerprinting. 
On two 5G datasets, our pre-trained \ac{TF} outperforms the localization accuracy of state-of-the-art methods when we subsequently fine-tune it on a small amount of data of the target environment for \ac{FP}-based localization.
Our self-pre-training approach achieves a similar localization accuracy to state-of-the-art supervised pre-training approaches, but requires ten times less reference data.
In the future, we will consider additional downstream tasks and different radio systems, investigate the generalization ability in dynamic environments, and investigate other architectures to reduce the computational effort. 

\section{Acknowledgment}\label{sec:acknowledgement}

This work was supported by the 
Federal Ministry of Education and Research of Germany in the programme of ``Souver\"an. Digital. Vernetzt.'' joint project 6G-RIC (16KISK020K),
the Fraunhofer Lighthouse project ``6G SENTINEL'' and by the Bavarian Ministry of Economic Affairs, Regional Development and Energy through the Center for Analytics – Data – Applications (ADA-Center) within the framework of ``BAYERN DIGITAL II" (20-3410-2-9-8).

\bibliography{library.bib}
\bibliographystyle{IEEEtran}

\end{document}